\documentclass[12pt,preprint]{emulateapj}

\shortauthors{ENOTO ET AL.}
\shorttitle{
Broad-band study with {\it Suzaku} of the magnetar class
}

\begin{document}
\title{
Broad-band study with {\it Suzaku} of the magnetar class
}


\author{
T. Enoto\altaffilmark{1,2}
K. Nakazawa\altaffilmark{1},
K. Makishima\altaffilmark{1,3},  
N. Rea\altaffilmark{4}, 
K. Hurley\altaffilmark{5},
and 
S. Shibata\altaffilmark{6}
}
\altaffiltext{1}{Department of Physics, University of Tokyo,
    7-3-1 Hongo, Bunkyo-ku, Tokyo, 113-0033, Japan;
     enoto@ceres.phys.s.u-tokyo.ac.jp}
\altaffiltext{2}{Current address: Kavli Institute for Particle Astrophysics and Cosmology, 
Department of Physics and SLAC National Accelerator Laboratory,  
Stanford University, Stanford, CA 94305, USA; enoto@stanford.edu} 
\altaffiltext{3}{High Energy Astrophysics Laboratory,
  Institute of Physical and Chemical Research (RIKEN),
    Wako, Saitama, 351-0198, Japan}
\altaffiltext{4}{Institut de Ciencies de l'Espai (ICE-CSIC, IEEC), Campus UAB, Facultat
de Ciencies, Torre C5-parell, 2a planta, 08193, Bellaterra
(Barcelona), Spain}
\altaffiltext{5}{Space Sciences Laboratory, 7 Gauss Way,
   University of California, Berkeley, CA 94720-7450, U.S.A.}
\altaffiltext{6}{Department of Physics, Yamagata University, Yamagata 990, Japan}

\begin{abstract}
Broad-band (0.8--70 keV) spectra 
	of the persistent X-ray emission
	from 9 magnetars 	
	were obtained with {\it Suzaku},
	including 3 objects in apparent outburst.
The soft X-ray component
	was detected from all of them,
	with a typical blackbody temperature of $kT\sim 0.5$ keV,
	while 
	the hard-tail component, dominating above $\sim$10 keV,
	was detected at $\sim$1 mCrab intensity from 7 of them.
Therefore,
	the spectrum composed of 
	a soft emission
	and
	a hard-tail component	
	may be considered to be a common property of magnetars,
	both in their active and quiescent states.
Wide-band spectral analyses
	revealed that 
	the hard-tail component
	has a 1--60 keV flux, $F_{\rm h}$,
	comparable to or even higher than that 
	carried by the 1--60 keV soft component, $F_{\rm s}$.
The hardness ratio of these objects,
	defined as $\xi \equiv F_{\rm h}/F_{\rm s}$,
	was found to be tightly anti-correlated with 
	their characteristic age $\tau_{\rm c}$ as 
	$\xi=(3.3\pm0.3)\times (\tau_{\rm c}/1 \textrm{\ kyr})^{-0.67\pm0.04}$
	with a correlation coefficient of $-0.989$,
	over the range from $\xi \sim 10$ to $\xi \sim 0.1$.
Magnetars in outburst states
	were found to lie on 
	the same correlation as relatively quiescent ones.
This hardness ratio is also positively correlated with 
	their surface magnetic fields with a correlation coefficient of $0.873$.
In addition,
	the hard-tail component 
	becomes harder 
	towards sources with older characteristic ages,
	with the photon index changing from $\sim$1.7 to $\sim$0.4.
\end{abstract}

	


\section{INTRODUCTION}
\label{introduction}

Soft Gamma Repeaters (SGRs) 
	and 
	Anomalous X-ray pulsars (AXPs)
	are considered to form 
	a distinct class of neutron stars,
	collectively called ``magnetars",
	and are thought to have magnetic fields 	
	exceeding the critical value of 
	$B_{\rm QED}=4.4\times 10^{13}$ G
	(\citealt{Thompson1995MNRAS,Thompson1996ApJ};
	for a recent review see, e.g., 
	\citealt{Mereghetti2008A&ARv,Woods2006csxs.book}).	
Although 
	their radiation,
	mainly emerging in the X-ray frequency,
	is considered to be powered by their huge magnetic energies,
	little is known about
	how the postulated strong magnetic fields 
	are dissipated and converted into the radiation.
It is hence imperative at this stage 
	to examine their emission properties 
	for any empirical relations 
	that provide a more unified characterization of these enigmatic objects.	

Through hard X-ray imaging observations
	with {\it INTEGRAL},
	some persistently bright magnetars 
	were discovered to emit 
	not only the well-known soft component
	(with a temperature of $kT\sim 0.5$ keV),
	but also 
	a distinct hard-tail component which emerges above $\sim$10 keV
	with an extremely hard photon index of 
	$\Gamma_{\rm h} \simeq 1$ (e.g. \citealt{Kuiper2006ApJ, Gotz2006A&A}).
The hard-tail component 
	has also been detected from 
	transient magnetars in outburst states,
	including 
	SGR~0501+4516 \citep{Rea2009MNRAS, Enoto2010ApJ}
	and 
	1E~1547.0$-$5408 \citep{Enoto2010PASJ}.
This unusual new component,
	though not yet observed from all magnetars,	
	is expected to provide a clue 
	to the emission process and the magnetic dissipation mechanism in magnetars.
	
To perform 
	wide-band spectroscopy of magnetars 
	covering these two components,
	multi-satellite observations have generally been employed.
However,
	the hard-tail intensity ($\lesssim$10 mCrab) of these objects
	requires much longer exposures
	than are usually spent in observing their soft components.
As a result,
	it has so far been difficult to obtain 
	strictly simultaneous wide-band spectra of these sources 
	which are more or less time-variable.
In the present {\it Letter},
	we report on {\it Suzaku} \citep{Mitsuda2007PASJ} observations 
	of most of the presently known magnetars, 
	studying for the first time soft and hard X-ray spectra taken 
	simultaneously. 
We thus found a strong correlation between magnetars' hardness ratio and characteristic ages.

\section{OBSERVATION AND ANALYSIS}
\label{obs_and_ana}

Since its launch,
	{\it Suzaku} has performed 
	15 pointed observations of 10 magnetars,
	including 5 targets in a {\it Suzaku} Key Project in 2009.
We analyzed all these {\it Suzaku} data
	in a unified way,
	except for 1E~1048.1$-$5937, 
	for which the data are not yet public  as of 2009 November.
This included reanalyses of 	
	the data from previously reported observations;
	i.e.,
	SGR~1806$-$20 \citep{Esposito2007A&A,Nakagawa2009PASJ},
	SGR1900+14 \citep{Nakagawa2009PASJ},
	1E~1841$-$045 \citep{Morii2008AIPC},
	CXOU J164710.2$-$455216 \citep{Naik2008PASJ},
	SGR~0501+4516 \citep{Enoto2009ApJL,Enoto2010ApJ},
	and 
	1E~1547.0$-$5408 \citep{Enoto2010PASJ}.
Table \ref{tab:ObsLog} summarizes the observations
	utilized in the present {\it Letter}.
Among them,
	three sources,
	namely,
	CXOU J164710.2$-$455216,
	SGR~0501+4516,
	and 
	1E~1547.0$-$5408,
	were observed during bursting periods.
Our sample, 
	though not complete,
	covers 9 sources out of the $\sim$15 magnetars known to date. 

From each observation,
	we extracted events for persistent X-ray emission using
	the X-ray Imaging Spectrometer (XIS; \citealt{Koyama2007PASJ})
	and
	the Hard X-ray Detector (HXD; \citealt{Takahashi2007PASJ}),
	which cover energy ranges of 
	0.2--12 keV and 10--600 keV, respectively.
The XIS and HXD data were processed 
	with the pipeline processing version 2.0 or later,
	and 
	events were discarded 
	if they were acquired
	in the South Atlantic Anomaly 
	or 
	in regions of low cutoff rigidity 
	($<$6 GV for the XIS and $<$8 GV for the HXD),
	or 
	at low Earth elevation angles\footnote{Detailed criteria are listed in; \\
	heasarc.nasa.gov/docs/suzaku/processing/criteria\_hxd.html \\
	heasarc.nasa.gov/docs/suzaku/processing/criteria\_xis.html
	}.
In addition,
	a fair number of short bursts, 
	detected from
	SGR~0501+4516, 1E~1547.0$-$5408, and SGR~1806$-$20,
	were removed 
	by discarding the XIS and HXD events 
	which were detected 
	within a time interval of a few seconds around each of these bursts
	(e.g., \citealt{Enoto2010PASJ}).
Studies of these burst events are beyond the scope of this {\it Letter}.	 	

To create 0.8--10 keV XIS spectra,
	we accumulated the screened data from 
	a source region typically within a 2$'$.0 radius
	of the target centroid,
	and 
	derived a background spectrum from 
	adjacent regions on the same CCD chip.
Since the background count rates
	are typically $\sim$5\% of the source rates,
	systematic uncertainties of the XIS background 
	are negligible.
Thanks to the use of appropriate window options,
	all the observations were free from XIS event pile up.	

For each observation,
	an HXD-PIN spectrum was produced in the 10--70 keV range,
	after subtracting 
	non X-ray background (NXB)
	and 
	Cosmic X-ray background (CXB; typically $\sim$4\% of the NXB).
We utilized NXB events
	which are simulated by the ``tuned" model
	of \cite{Fukazawa2009PASJ},
	and 
	the analytic CXB model 
	by \cite{Moretti2009A&A}.
In the case of sources near the Galactic center
	(i.e., SGR~1806$-$20 and 1RXS~J170849.0$-$400910), 
	we further subtracted 
	Galactic Ridge X-ray Emission 
	(GRXE, \citealt{Krivonos2007A&A}; typically $\lesssim$3\% level of the NXB),
	with a photon index fixed at 2.1 \citep{Valinia1998ApJ}
	and 
	normalizations 
	specified by observations of nearby blank skies at similar Galactic latitudes.
Systematic errors in these HXD-PIN spectra
	are dominated by 	
	reproducibility of the NXB model,
	which is typically $\lesssim$2\% (1$\sigma$) 
	\citep{Fukazawa2009PASJ},
	while 
	those of the CXB and GRXE are negligible.
This effect is taken into account
	by changing the PIN-NXB within $\pm$2\%.
The HXD-PIN field-of-view was clear of significant contaminating sources 
	in these observations,
	except for the case of 	
	CXOU J164710.2$-$455216 \citep{Naik2008PASJ},
	which was confused with GX~340+0.
Source confusion was not an issue for the other 8 sources,
	and we hence analyze these 8 sources below.

\section{RESULTS}
\label{results}

Figure \ref{fig:fig1} shows
	background-subtracted and pulse- phase-averaged persistent spectra 
        of the 8 magnetars.
The soft component
	was clearly observed from all of them,
	while
	the hard-tail component
	was successfully detected from 
	7 of them at a typical intensity of $\sim$1 mCrab (at $\sim$30 keV).
Assuming $\Gamma_{\rm h} \sim 1$,
	any hard-tail flux from 1E~2259+586 
	was constrained to be less than 
	$8.7\times 10^{-12}$ erg\,s$^{-1}$\,cm$^{-2}$
	in the 15--60 keV range,
	which is consistent with the previous result 
	from {\it INTEGRAL}
	(Figure~10 in \citealt{Kuiper2006ApJ}).
These {\it Suzaku} spectra reconfirm the previous {\it INTEGRAL} 
	detections of the hard-tail component
	from some persistent sources,
	including 	
	4U~0142+61,
	1RXS~J170849.0$-$400910,
	1E~1841$-$045,
	SGR~1806$-$20,
	and 
	SGR~1900+14
\citep{Mereghetti2005A&A,Kuiper2006ApJ,Gotz2006A&A,Hartog2007Ap&SS,Hartog2008A&A}.
The absorbed 2--10 keV and 15--60 keV X-ray fluxes 
	are summarized in Table \ref{tab:ObsLog}.
Thus,
	the hard-band flux is comparable to,
	or sometimes even higher than,
	that in the softer band.	

The unified presentation in Figure~\ref{fig:fig1}
	yields an intriguing inference 
	that 
	the hard-tail component
	becomes weaker
	for older objects.
To quantify this property,
	we simultaneously fitted the XIS and the HXD spectra of each object with
	a common spectral model.
For the hard-tail component,
	we employed a single power-law for all the sources.
The soft component of younger magnetars 
	was reproduced 
	by a single blackbody (Model A),
	but 
	older sources required some modification to this model
	due to a ``soft-tail" excess around 8 keV,
	which is thought to arise via some sort of Comptonzation process 
	\citep{Thompson2002ApJ, Lyutikov2006MNRAS, Rea2008ApJ}.
We therefore introduced 
	an empirical Comptonized blackbody model 
	(Model B; \citealt{Tiengo2005A&A,Halpern2008ApJ,Enoto2010PASJ}),
	and obtained successful fits in most cases.
In the case of SGR~0501+4516,
	neither Model A nor B 
	was acceptable, but 
	a successful fit was obtained by adding a cooler blackbody to Model B
	(Model C; \citealt{Enoto2010ApJ}).
The XIS spectrum of 	1E 1841-045
	needed an additional plasma emission component 
	from the surrounding supernova remnant.

From these fit results, 
	we calculated 
	absorption-corrected fluxes for
	the soft component $F_{\rm s}$ 	
	and 
	the hard-tail component $F_{\rm h}$,
	both defined in the 1--60 keV range.
These values, 
	given in Table~\ref{tab:ObsLog},
	are considered physically more meaningful 
	than the raw 2--10 keV and 15--60 keV fluxes,
	because the two components are separated,
	and the absorption is corrected.
The lower boundary of 1 keV was adopted 
	not only for $F_{\rm s}$ but also for $F_{\rm h}$, 
	because sources with young characteristic ages 
	have substantial contributions of the hard-tail
	even in the lowest energy range at $\sim$0.8 keV (e.g., SGR 1806$-$20 in Figure~\ref{fig:fig1}).
We may then define the ``hardness ratio" (HR),
	$\xi \equiv F_{\rm h}/F_{\rm s}$.
As shown in Table~\ref{tab:ObsLog},
	the values of $\xi$ range over almost two orders of magnitude from 0.2 to 10.
	
As visualized in Figure~\ref{fig:fig2},
	$\xi$ exhibits a clear gradient on the $P$-$\dot{P}$ plane.
To assess this implication,
	we show in Figure~\ref{fig:fig3}
	the values of $\xi$ 
	as a function of 
	the characteristic age
	$\tau_{\rm c}=P/2\dot{P}$\,s
	and
	the surface magnetic field 	
	$B_{\rm s}=3.2\times 10^{19} (P\dot{P})^{1/2}$\,G,
	where the errors associated with $\xi$ 
	include both statistical and systematic errors.
The latter is dominated by 
	the 2\% uncertainties of PIN-NXB (\S~\ref{obs_and_ana}),
	with additional contributions (by $\sim$1\% in $\xi$)
	from uncertainties in the XIS vs HXD cross-normalization.
The figure indicates that 	
	$\xi$ is indeed 
	negatively (or positively)
	correlated 
	with $\tau_{\rm c}$ (or $B_{\rm s}$).
	
	
After discarding the two upper limits in Figure~\ref{fig:fig3},
	$\xi$ is found to be correlated with $\tau_{\rm c}$ and $B_{\rm s}$
	with Pearson's linear cross-correlation coefficients
	of $r = -0.989$ and $0.873$, respectively.
Thus, $\xi$ is very tightly correlated with $\tau_{\rm c}$,
	and to a lesser extent (as evident in Figure~\ref{fig:fig2}), with $B_{\rm s}$.	
In order to evaluate the correlation in a non-parametric way,
	we further performed the Spearman's rank-order test,
	which gives correlation coefficients of 
	$r_s = -0.963$ and $0.871$
	for $\tau_{\rm c}$ and $B_{\rm s}$, respectively,
	together with a chance probability in both cases of $<$0.001.
Thus, these correlations are indeed significant.
Fitting through the least squares method
	gave the best-fit linear correlation as 
\begin{eqnarray}
\xi=\frac{F_{\rm h}}{F_{\rm s}}&=&
(3.3\pm0.3)\times 
\left(\frac{\tau_{\rm c}}{1 \textrm{\ kyr}}\right)^{-0.67\pm0.04}
\label{Eq:HR_tauc}
\\
&=&
(0.09\pm 0.07) \times 
\left(\frac{B_{\rm s}}{B_{\rm QED}}\right)^{1.2\pm 0.2},
\label{Eq:HR_Bs}
\end{eqnarray}
with 1$\sigma$ uncertainties.
Since individual data points 
	have rather small errors,
	we here added constant systematic errors to them,
	so as to make the fits acceptable
	with reduced chi-square values of $\sim$1.
Around the $\xi$-$\tau_{\rm c}$ linear fit,
	the data points scatter by no more than a factor of 2 in $\xi$.

Now that the $\xi$-$\tau_{\rm c}$ and the $\xi$-$B_{\rm s}$ correlations
	were confirmed to be statistically significant, 
	let us examine them for systematic effects.
These correlations are unlikely to arise from some selection biases on the $P$-$\dot{P}$ plane,
	because these magnetars 
	show a 2-dimensional scatter in Figure~\ref{fig:fig2} without any particular correlation
	between $P$ and $\dot{P}$.	
These observations are also free from possible uncertainties 
	due to independent variations of $F_{\rm s}$ and $F_{\rm h}$,
	since in all sources
	these two quantities are based on strictly simultaneous measurements.
In addition,	
	multiple observations of the same objects
	at different epochs
	gave similar values of $\xi$
	(i.e., SGR~1806$-$20,
	SGR~1900+14,
	and 4U~0142+61 in Figure~\ref{fig:fig3}).
Therefore,
	the essential features of Figure~\ref{fig:fig2} are not 
	considered to depend strongly on the epochs of observations.
	
Although 
	the empirical Comptonized blackbody model (Model B)
	was employed in calculating $F_{\rm s}$ (and hence $\xi$) 
	of 
	1RXS~J170849.0$-$400910, SGR~0501+45, and 4U~0142+61,
	their values of $\xi$ change only by a factor of 
	1.78, 1.09, and 0.94--1.09, respectively,
	if their soft component is modeled by two blackbodies 
	that constitute another popular model.
Thus, the implication of Figure~\ref{fig:fig3} remain unchanged.
The $\xi$-$\tau_{\rm c}$ correlation also remains robust, 
	even if we change the lower limit of the energy band 
	from 1 keV to 0.5 keV. 
For example,
	this change reduces $\xi$ by only 5\% and 26\% 
	for the cases of SGR 1806$-$20 and 4U 0142+61, respectively. 
Similarly,
	if we raise the upper-limit energy of the hard-tail flux 
	from 60 keV to 500 keV,
	the $\xi$-$\tau_{\rm c}$ correlation also persists 
	with $r=-0.788$ and $r_s=-0.825$.
Therefore,	
	the $\xi$-$\tau_{\rm c}$ relation can be considered to represent 
	a model-independent property of our sample of magnetars.
Detailed spectral analyses of our sample objects
	will be described elsewhere,
	together with separate studies of 
	pulsed and un-pulsed signals.

\section{DISCUSSION AND CONCLUSION}
\label{discussion_and_conclusion}

\begin{figure*}[h]
\begin{center}
\includegraphics[scale=0.8]{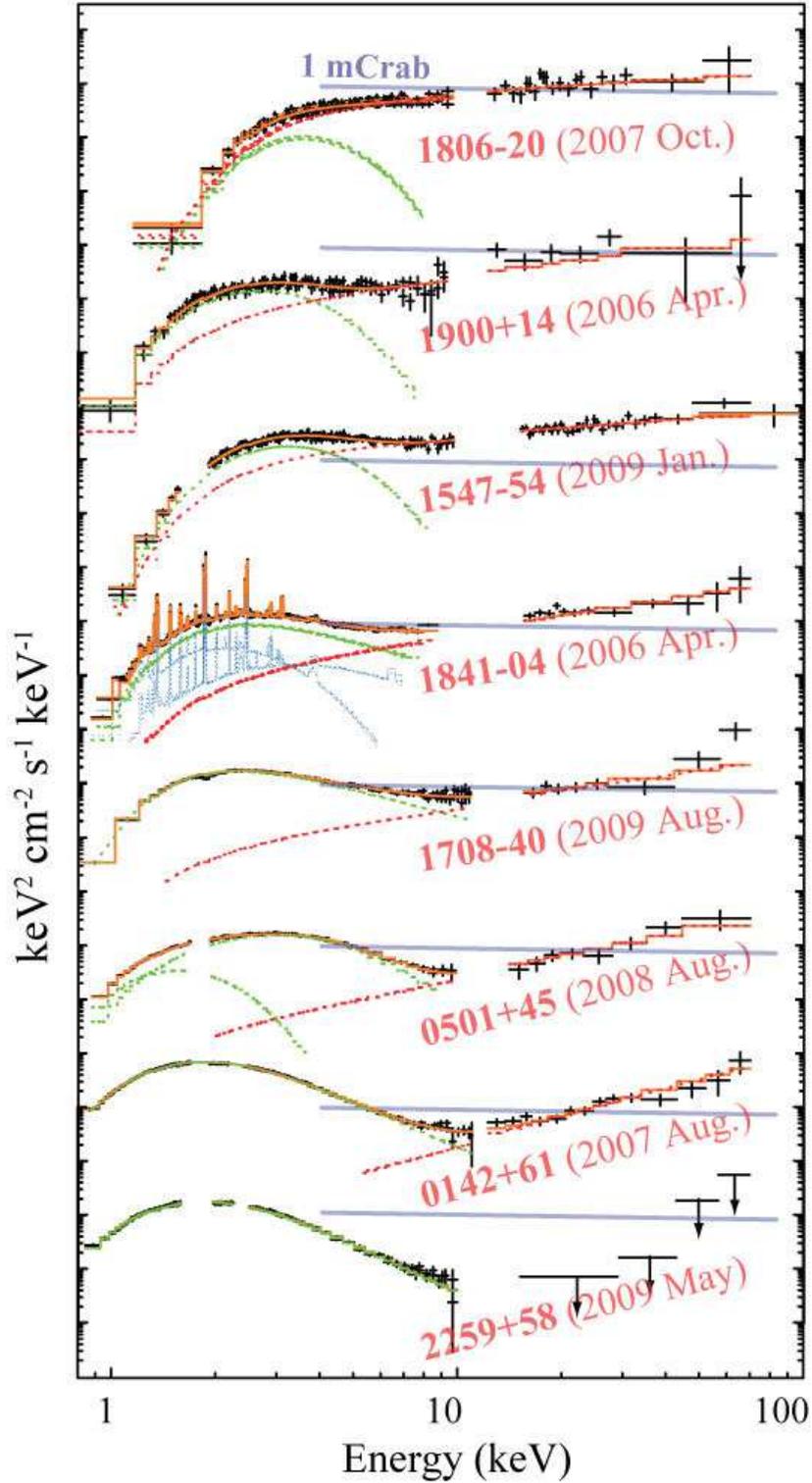}
\caption{
Background-subtracted $\nu F_{\nu}$ spectra of 
	the persistent emission of the magnetars,
	shown after eliminating the instrumental responses.
Interstellar absorption is included.
Individual spectra are shown with offsets,
	and
	are arranged in order of increasing characteristic age
	from top (young) to bottom (old).
Blue horizontal lines indicate
	a 1 mCrab intensity.
Green, red, and blue lines
	represents
	the soft component,
	the hard-tail component,
	and 
	the SNR contamination (including line emission) in the 1E~1841-045 spectrum,
	respectively.
If a source was observed more than once,
	one observation is shown.
The GSO data of 1E~1547.0$-$5408 are included after \cite{Enoto2010PASJ}.
The full names of the sources,
	which are shown here as abbreviations,
	are listed in a footnote to Table~\ref{table}.
}
\label{fig:fig1}
\end{center}
\end{figure*}

\begin{figure*}[h]
\begin{center}
\includegraphics[scale=0.9]{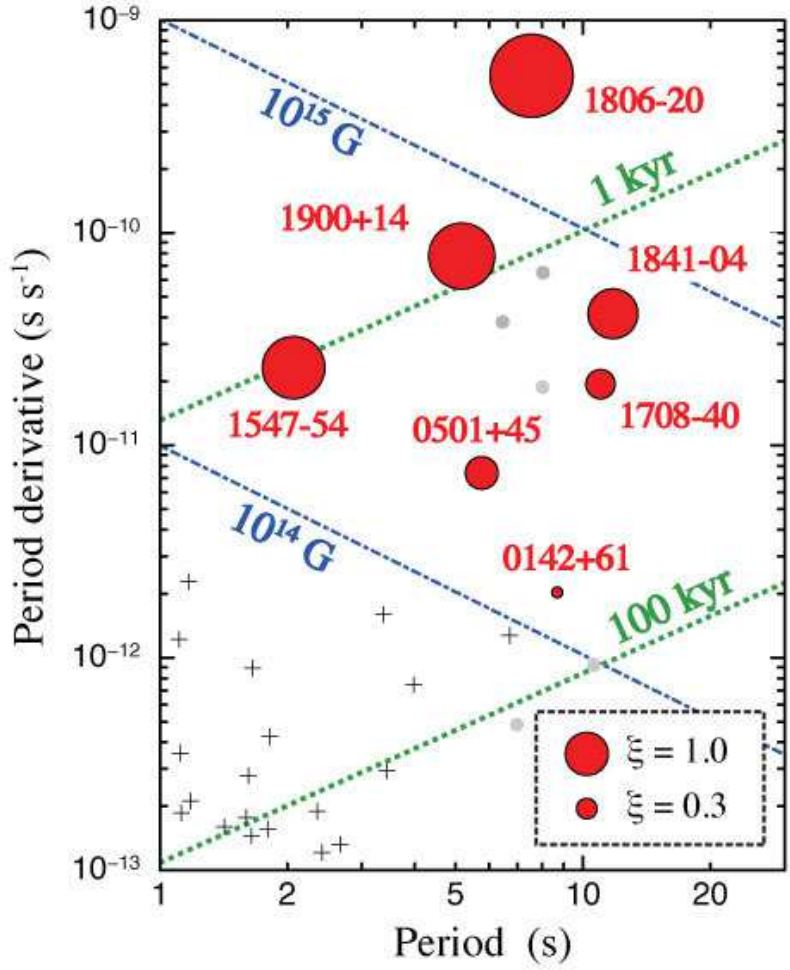}
\caption{
A portion of the $P$-$\dot{P}$ diagram of pulsars
	(\citealt{Manchester2005}; http://www.atnf.csiro.au/research/pulsar/psrcat/).
Sizes of the red circles 
	correspond to the $\xi$ values
	in the present sample.
Blue and green lines represent
	constant values of 
	the magnetic field strength
	and 
	the characteristic age, 
	respectively.
Small grey circles represent
	magnetars with upper limits on the hard-tail component,
	or those not observed with {\it Suzaku}.
Ordinary radio pulsars are indicated by crosses.	
}
\label{fig:fig2}
\end{center}
\end{figure*}

\begin{figure*}[h]
\begin{center}
\includegraphics[scale=0.8]{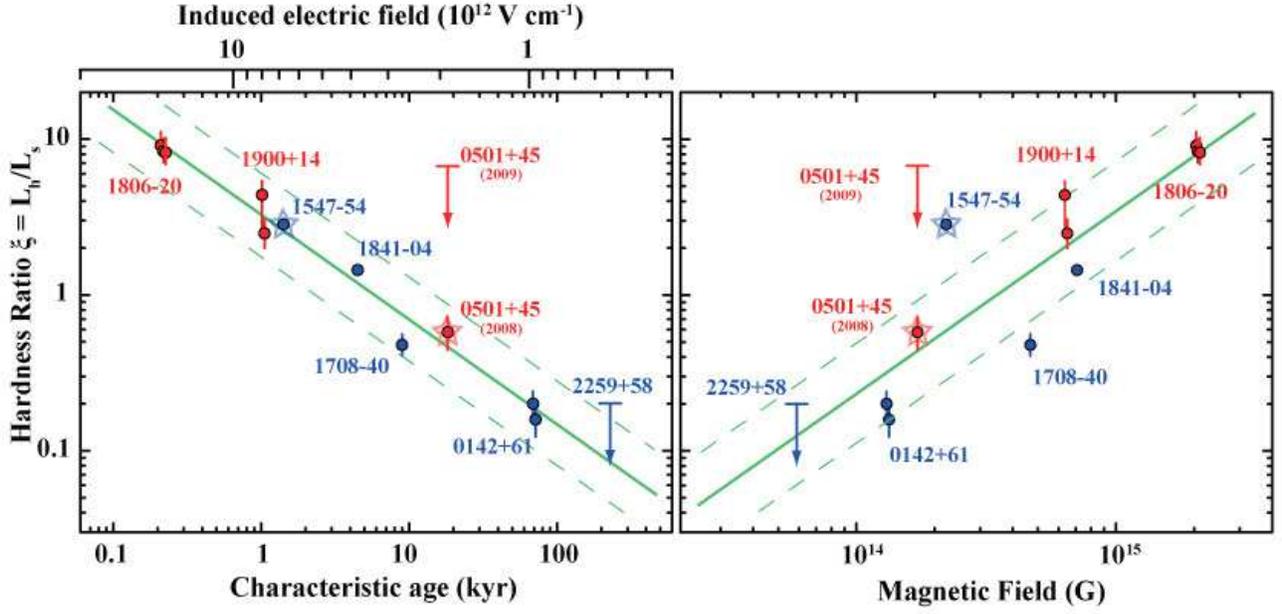}
\caption{
(left) 
A correlation between
	the HR $\xi$ and the characteristic age $\tau_{\rm c}$.
Quoted errors are statistical plus systematic 1$\sigma$ (Table~\ref{tab:ObsLog}).
Upper limits are also shown for SGR\,0501+4516 
	(2nd observation) and 1E\,2259+586.
Green solid and dashed lines represent
	the best fit of equation (\ref{Eq:HR_tauc})
	and their boundaries shifted by a factor of two, respectively.
SGRs and AXPs are shown in red and blue, respectively,
	while burst-active sources are shown with star symbols.
(Right)
The same as the left panel,
	but plotted as a function of the surface magnetic field.
}
\label{fig:fig3}
\end{center}
\end{figure*}

Including most of the known SGRs and AXPs,
	the present data set provides 
	the first simultaneous coverage of their 
	soft and hard X-ray spectra (Figure~\ref{fig:fig1}).
This has led to a discovery of the 
	$\xi$-$\tau_{\rm c}$ and $\xi$-$B_{\rm s}$ correlations in Figure~\ref{fig:fig3}.
These tight correlations suggest
	that 
	the values of $\xi$
	in the present objects 
	approximately behave as a one parameter family
	(a single valued function)
	determined uniquely by $P$ and $\dot{P}$.
Among possible physical quantities 
	expressed as simple combinations of $P$ and $\dot{P}$,
	the characteristic age $\tau_{\rm c}$ 
	may be regarded as a likely determinant of $\xi$
	(see Figure~\ref{fig:fig2}).
Then, by regarding $\tau_{\rm c}$ 
	as an indicator of the age in assuming a self-similar spin-down,
	and 
	further noting that 
	our sample covers a significant ($\sim$50\%) fraction of the presently 
	catalogued objects of these two classes,
	the $\xi$-$\tau_{\rm c}$ correlation is considered to represent
	a common spectral evolution of them.
Since both SGRs and AXPs follow
	the same correlations,
	the present result
	provides another support to the increasing evidence 
	that these subsets 
	are intrinsically considered the same kinds of object, namely ``magnetars".

From the burst-active magnetars SGR~0501+4516 and 1E~1547.0-5408,
	we successfully detected the hard-tail components.
On the $\xi$-$\tau_{\rm c}$ correlation,
	their $\xi$ values generally agree with 
	data points of the other magnetars,
	which mainly refer to quiescent or less active states.
Therefore, 
	the bursting activity of a magnetar 
	is thought to cause
	a significant enhancement in the hard component, 
	roughly to the same intensity
	as the soft component,
	which is known to increase by 1--2 orders of magnitude 
	\citep{Tam2008ApJ,Rea2009MNRAS,Enoto2010PASJ}.
The upper limit on the 2nd observation of SGR~0501+4516
	is consistent with this picture.
These results further suggest that 
	the basic mechanism of quiescent wide-band emission 
	is retained during increased activity.

	
In addition to 
	the above interpretation invoking $\tau_{\rm c}$,
	the same magnetar $\xi$-$\tau_{\rm c}$ correlation
	may be interpreted in an alternative way.
The strong magnetic field $B$,
	combined with the rotation,
	would induce 
	an electric field up to
\begin{eqnarray}
|\vec{E}| &\sim& R\Omega B 
	\nonumber \\ 
	&=&
	2.6 \times 10^{12} \  \frac{{\rm V}}{{\rm cm}}
	\left(	\frac{6 \textrm{\ s}}{P} \cdot 
	\frac{\dot{P}}{10^{11} \textrm{\ s\,s$^{-1}$}} \right)^{1/2}
	\left( \frac{R}{10 \textrm{\ km}} \right)
	\nonumber \\
	&\propto& \sqrt{\frac{1}{\tau_{\rm c}}}
	\ ,
\end{eqnarray}
where $R$ and $\Omega$ are 
	the distance from the center of the star
	and the angular frequency,
	respectively.
Thus, 
	$\tau_{\rm c}$ can be uniquely converted to $|\vec{E}|$,
	as shown at the top of Figure~\ref{fig:fig3} (left).
Thus,
	the magnetar $\xi$-$\tau_{\rm c}$ correlation 
	can also be regarded as 
	a correlation between $\xi$ and $|\vec{E}|$.
If so, 
	the hard-tail component,
	which should require the presence of particles
	more energetic than $\sim$100 keV,
	might be related to particle acceleration by this induced electric field.

\begin{figure*}[h]
\begin{center}
\includegraphics[scale=0.6]{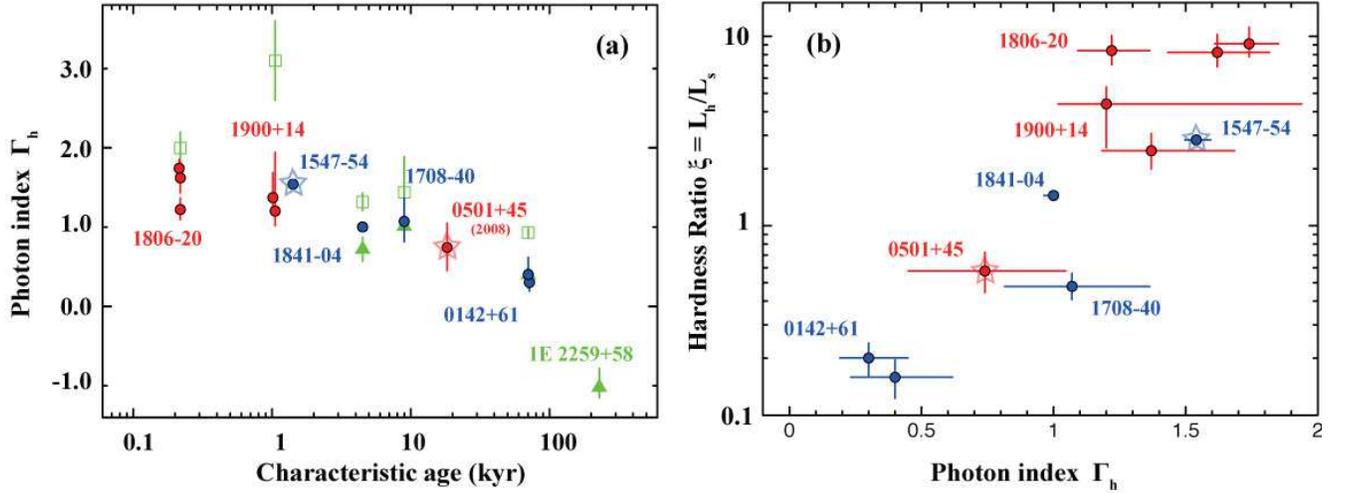}
\caption{
(a) Photon indices $\Gamma_{\rm h}$ of the hard-tail component of the present sample,
	plotted as a function of the characteristic age.
Quoted errors are statistical and systematic 1$\sigma$.
Colors and symbols are the same as in Figure~\ref{fig:fig3}.
Green squares and triangles 
	represent photon indexes of pulsed and total fluxes,
	respectively, after \citet{Kaspi2010ApJL}.
(b) Relation between $\Gamma_{\rm h}$ and $\xi$.	
}
\label{fig:fig4}
\end{center}
\end{figure*}

A closer inspection of Figure~\ref{fig:fig1}
	suggests
	a hardening of the hard-tail component
	for objects with larger $\tau_{\rm c}$.
This is more clearly shown in Figure~\ref{fig:fig4}(a),
	where 
	$\Gamma_{\rm h}$ is observed to evolve 
	from $\sim$1.7 to $\sim$0.4.
Such a correlation was first pointed out in the 1--10 keV band
	by \cite{Marsden2001ApJ},
	and 
	recently 
	extended 
	into hard X-rays by \cite{Kaspi2010ApJL}.
The latter authors found a correlation
	of $\Gamma_{\rm h}$ with $B_{\rm s}$ 
	or $\dot{P}$,
	with implications similar to those of our Figure~\ref{fig:fig4}(a).
In contrast,
	they did not find any correlations that 
	can be compared with our Figure~\ref{fig:fig3},
	presumably due to lack of simultaneity 
	between  the soft and hard X-ray measurements.	
Thanks to the simultaneous observations with {\it Suzaku},
	we can plot, in Figure\ref{fig:fig4}(b),
	the values of $\xi$ 
	as a function of $\Gamma_{\rm h}$.
Contrary to normal expectations 
	that 
	sources with lower broad-band hardness ratios
	should have softer (larger) values of $\Gamma_{\rm h}$,
	this figure	reveals an opposite correlation,
	that older magnetars with weaker hard-tail fluxes (smaller $\xi$)
	in fact have harder (smaller) slopes $\Gamma_{\rm h}$.
Since this is not affected by the selection of energy bands (\S3),
	such a trend should imply an intrinsic property 
	of the hard-tail production mechanism.

The present magnetar observations
	require 
	the hard-tail emission mechanism 
	to satisfy 
	both the $\xi$-$\tau_{\rm c}$ correlation
	and 
	the hardening of the hard-tail component towards older objects.
The difference of $\Gamma_{\rm h}$ among the sources 
	rules out 
	those models which predict 
	a fixed common hard-tail slope 
	(e.g., the fast-mode breakdown model; \citealt{Heyl2005MNRAS}).
Such a large variation of $\Gamma_{\rm h}$ with $\tau_{\rm c}$
	is not expected by  
	a thermal bremsstrahlung model from 
	the transition layer on the stellar surface, either
	\citep{Thompson2005ApJ,Beloborodov2007Ap&SS}.
Similarly, observed $\Gamma_{\rm h}$ values have not yet been reproduced 
	by existing resonant magnetic Compton up-scattering models
	\citep{Baring2007Ap&SS,Fernandez2007ApJ}.
It remains to be examined 
	whether the results of the present observations  
	can be explained by
	any of the existing (or future) theoretical explanations
	(e.g., those based on the ``twisted magnetosphere" model; \citealt{Thompson2002ApJ}).
	
As an altenative possibility,
	a photon-splitting process can act to produce the hard-tail emission.
At the neutron-star surface, 
	sub-MeV photons can be produced 
	through electron-positron annihilation or
	resonant magnetic Compton up-scatterings.
Since there is no low energy threshold for the photon splitting \citep{Harding2006},
	these sub-MeV photons may repeat splitting \citep{Harding1997ApJ, Baring2001ApJ}
	in the strong field, 
	to form the hard continuum downward towards lower energies.
In this case,
	the higher magnetic fields of younger magnetars
	will allow the photons to repeat splitting toward lower energies,
	thus making $\Gamma_{\rm h}$ larger in agreement with our figure.

\acknowledgements 
We thank members of the {\it Suzaku} magnetar Key Project,
	including Y. E. Nakagawa and M. Morii 
	for discussion on SGR~1900+14 and 1E~2259+586.
We are also grateful to L. Dong, N. Shibazaki, and M. Baring 
	for suggestions on the $\xi$-$\tau_{\rm c}$ correlation.


\newpage

\begin{table}
\begin{center}
\caption{Log of {\it Suzaku} observations of magnetars.\label{tab:ObsLog}}
\label{table}
\scriptsize
\begin{tabular}{p{0.9cm}p{0.5cm}p{0.4cm}p{0.18cm}p{0.1cm}p{1.05cm}p{1.1cm}p{0.4cm}p{1.6cm}p{1.7cm}p{2.3cm}p{1.2cm}p{0.7cm}}
\tableline 
\tableline
Name\tablenotemark{a}  &
\multicolumn{1}{c}{$P$\tablenotemark{b}}  & 
\multicolumn{1}{c}{$\dot{P}$\tablenotemark{b}} & 
\multicolumn{1}{c}{$B_{\rm s}$\tablenotemark{c}} &
\multicolumn{1}{c}{$\tau_{\rm c}$\tablenotemark{d}}  &
\multicolumn{1}{c}{Obs. ID} & 
\multicolumn{1}{c}{Date}   & 
Exp.\tablenotemark{e}  & 
\multicolumn{1}{c}{Model\tablenotemark{f}}  &
\multicolumn{1}{c}{Abs. $F_{\rm x}$\tablenotemark{g}}  &
\multicolumn{1}{c}{Unabs. $F_{\rm x}$\tablenotemark{h}} & 
\multicolumn{1}{c}{HR\tablenotemark{i}} &
\multicolumn{1}{c}{$\Gamma_{\rm h}$\tablenotemark{j}} 
  \\
	&  & $10^{-11}$ & $10^{14}$  & &  & &    &                                &   \multicolumn{1}{c}{$F_{15-60}$/$F_{2-10}$} &
	\multicolumn{1}{c}{$F_{\rm h}$/$F_{\rm s}$} &  $\xi$$=$$F_{\rm h}/F_{\rm s}$ \\
       	& (s) & {\tiny (s\,s$^{-1}$)} & {\tiny (G)} & {\tiny (kyr)}
	&   &  {\tiny yymmdd}	 & (ks)        
	&   \multicolumn{1}{c}{($\chi^2$/d.o.f)}
	& \multicolumn{1}{c}{15--60/2--10} & \multicolumn{1}{c}{1--60/1--60} &  \\
\tableline		   
\multicolumn{12}{c}{Soft Gamma-ray Repeater}\\    
1806$-$20 & 7.556 & 54.9 & 21 & 0.22 &  401092010 & 06-09-09   & 55.4
		   & A(468.9/434) &30.2(3)/12.6(1)   
		   & $55.6^{+3.1}_{-2.3}$/$6.1^{+1.3}_{-0.9}$ & $9.1^{+2.1}_{-1.3}$  
		   & 1.7(1) \\
                     & $\cdots$ & $\cdots$ & $\cdots$ & $\cdots$ & 401021010  & 07-03-30    & 16.5 &
                     A(185.1/182) & 46(10)/9.7(3) & $57.1^{+6.5}_{-6.0}$/$6.8^{+1.1}_{-0.8}$  &
                      $8.4^{+1.7}_{-1.3}$ & 1.2(1) \\
                     & $\cdots$  & $\cdots$  & $\cdots$ & $\cdots$ &  402094010  & 07-10-14   & 50.1 & 
                     A(239.3/240) & 24(3)/8.6(2) & $38.6^{+4.0}_{-7.7}$/$4.7^{+1.0}_{-0.7}$ & 
                     $8.2^{+2.0}_{-1.3}$  & 1.6(2) \\
1900+14 & 5.169 & 7.78 & 6.4 & 1.1 &  401022010& 06-04-01  & 14.4 
		   & A(145.5/156) & 15(6)/4.5(3) & $18.7^{+4.0}_{-7.7}$/$4.3^{+0.3}_{-0.2}$ & 
		   $4.4^{+1.0}_{-1.8}$  & 1.2(5) \\
		 & $\cdots$  & $\cdots$ & $\cdots$ & $\cdots$ &   
		404077010& 09-04-26   & 42.1 
		 & A(433.6/393) & 8(4)/3.2(2) & $10.7^{+2.4}_{-2.1}$/4.3$\pm 0.2$ & 
		 $2.5^{+0.6}_{-0.5}$ & 1.4(3) \\
0501+45    & 5.762 & 0.5 & 1.7 & 18  & 903002010 & 08-08-26   & 33.5 &
		C(151.2/149) & 27(6)/27.2(5) & $27.4^{+6.9}_{-6.3}$/$47.5^{+1.0}_{-1.4}$ &
		$0.58^{+0.15}_{-0.13}$ & 0.7(3)  \\
		%
                     & $\cdots$ & $\cdots$ & $\cdots$ & $\cdots$ &  404078010  & 09-08-17   & 43.0
                     & B(191.7/154) & $<$24/1.4(4)  & $<$25/3.8$\pm$0.2 & $<$6.7 
                     &  --- \\
\tableline
\multicolumn{12}{c}{Anomalous X-ray Pulsars} \\
1547$-$54 & 2.070 & 2.32 & 2.2 & 1.4 &  903006010 & 09-01-28    & 33.5 &
		      A(298.9/278)  
		      & 105(7)/57(1) & 159$\pm$5/$56^{+1}_{-2}$ & 2.8$\pm$0.1&
		      1.54(5)\\
1841$-$04  & 11.775  & 4.16 & 7.1 & 4.5 &  401100010 & 06-04-19 & 63.5 &
		     B(462.9/262) & 45(4)/26(2) & $49.7^{+1.6}_{-1.1}$/$34.4^{+0.5}_{-0.3}$ & $1.44^{+0.05}_{-0.04}$  & 1.00(3)\\
1708$-$40 &  10.999 & 1.94 & 4.7& 9.0 &  404080010 & 09-08-23 & 60.9  &
		     B(94.6/98)& 27(6)/28.8(6) & $30.0^{+5.1}_{-4.2}$/$62.7^{+2.7}_{-3.0}$ & $0.48^{+0.08}_{-0.07}$  & 1.1(3)  \\
0142+61     & 8.688  & 0.20 & 1.3 & 70 &  402013010 & 07-08-13     & 101.6 &
		   B(434.4/408) & 36(3)/64.4(2) & $36.3^{+7.1}_{-7.0}$/$181.1^{+0.5}_{-1.2}$ & 0.20$\pm$0.04 & 0.3(1) \\
                     & $\cdots$ & $\cdots$ & $\cdots$ & $\cdots$ &  404079010 & 09-08-12     & 107.4  &
                      B(310.6/301)  & 30(4)/61.4(3) & $27.8^{+6.4}_{-6.3}/175.2^{+1.0}_{-1.2}$ & 0.16$\pm$0.04 & 0.4(2) \\
1647$-$45 & 10.611 & 0.24 & 1.6 & 70  &  901002010 & 06-09-23 & 35.1 &
		   B(333.2/294) & ---/23.4(3)  & ---/45.1$\pm$0.5 & ---  & ---\\
2259+58     & 6.979 & 0.05 & 0.59& 230 &  404076010 & 09-05-25     & 103.4  & 
		  B(222.1/213) & $<$8.7/15.8(1) & $<$9.2/45.7$\pm$0.4 & $<$0.2 & --- \\
\tableline
\end{tabular}
\tablenotetext{a}{Abbreviated names
	represent 
	SGR~1900+14, 
	SGR~1806$-$20,
	SGR~0501+4516,
	1E~1547.0$-$5408,
	1E~2259+586,
	4U~0142+61,
	CXO~J164710.2$-$455216,
	1RXS~J170849.0$-$400910,
	and 
	1E~1841-045.}
\tablenotetext{b}{$P$ and $\dot{P}$ values are from the 
	McGill SGR/AXP Online Catalog (http://www.physics.mcgill.ca/$\sim$pulsar/magnetar/main.html).}
\tablenotetext{c}{Surface magnetic field strength estimated as 
	$B_{\rm s}=3.2\times 10^{19} (P \dot{P})^{1/2}$\,G.}
\tablenotetext{d}{Characteristic ages estimated as 
	$\tau_{\rm c}=P/2\dot{P}$\,s.}
\tablenotetext{e}{Effective Exposures of the HXD.}
\tablenotetext{f}{Model A: a blackbody plus a power-law model (BB+PL);
	Model B: a Comptonized blackbody plus the power-law (CBB+PL);
	Model C: a blackbody  and the Comptonized blackbody plus the power-law (BB+CBB+PL).
	}
\tablenotetext{g}{Absorbed X-ray fluxes ($10^{-12}$\,erg\,s$^{-1}$\,cm$^{-2}$) 
	in the 15--60 keV and 2--10 keV bands. 
	Quoted errors are statistical-only at the 90\% confidence level. }
\tablenotetext{h}{Unabsorbed X-ray fluxes ($10^{-12}$\,erg\,s$^{-1}$\,cm$^{-2}$) 
	of the 1--60 keV soft component and the 1--60 keV hard-tail component.
	Statistical errors (1$\sigma$) are included. 
	Systematic errors (1$\sigma$) are also included in the hard-tail component,
	calculated from the uncertainties in the PIN-NXB
	and in the cross-normalization between the XIS and HXD-PIN. 
	}
\tablenotetext{i}{Hardness ratio $\xi$, calculated from the two unabsorbed fluxes.
	Quoted errors are statistical and systematic 1$\sigma$ level.}
\tablenotetext{j}{Photon index of the hard-tail component. Quoted errors are statistical and systematic 1$\sigma$ level. If the upper and lower errors are different, average errors are shown.}
\end{center}
\end{table}

\clearpage

\end{document}